\documentstyle[aps,prb,multicol,epsf,picinpar]{revtex}

\def\ep {\epsilon}

\def\e2 {\epsilon-\epsilon_k}
\def\be {\begin{equation}}
\def\ee {\end{equation}}
\def\bea {\begin{eqnarray}}
\def\eea {\end{eqnarray}}
\def\om {\omega}

\begin{document}
\draft
\title{A Fermi liquid model for the overdoped and optimally doped 
cuprate superconductors: scattering rate, susceptibility,
spin resonance peak and superconducting transition}

\bigskip
\author{George Kastrinakis}
\address{Dept. of Physics, University of Illinois at Urbana-Champaign,
Urbana, IL 61801, USA\\
and\\
Dept. of Chemical Engineering, University of Cambridge,
Cambridge CB2 3RA, U.K. $^*$ 
 }

\date{Received 14 Jan. 2000}
\author{Physica C {\bf 340}, 119 (2000)}

\maketitle
\begin{abstract}
We present a Fermi liquid model
for the overdoped and optimally doped cuprate superconductors.
For the normal state,
we provide an analytic demonstration, backed by
self-consistent Baym-Kadanoff (BK) numerical calculations, of the linear in 
temperature
resistivity and linear in 1/energy optical conductivity,
provided the interacting Fermi liquid has strong peaks in its density
of states (van-Hove singularities in 2 dimensions) near the chemical potential
$\mu$.
Recent ARPES expts. by Valla et al., Science {\bf 285}, 2110 (1999), 
and e-print cond-mat/0003407, 
directly support the linearity of the one-particle scattering rate
everywhere in the Brillouin zone hereto obtained.
We show that the origin of this linearity is the linear in energy term
of the imaginary part of the carrier susceptibility.
Moreover, we verify that the interactions tend to pin the
van-Hove singularities close to $\mu$.
We show that the {\em low} energy dependence of the 
susceptibility 
can have a purely fermionic origin.
We introduce an {\em ansatz} for the susceptibility of the carriers, which
we postulate to be {\em enhanced} in an additive
manner due to the weak antiferromagnetic order of the CuO$_2$ planes.
Inter alia, this ansatz may explain the appearance of the spin resonance peak
(observed in neutron scattering) in the normal state of the cuprates. 
Further, we obtain particularly high transition temperatures $T_c$ from our
BK-Eliashberg scheme
by using this ansatz:
we have a $d_{x^2-y^2}$ gap with $T_c > 120 ^o$K for
nearest neighbour hopping $t=250meV$.
                   
\end{abstract}


\vspace{1cm}
\centerline{\bf I. Introduction}
\vspace{1cm}

The nature of the many-body state of the cuprate superconductors
is a central question for the understanding of these materials
\cite{agl,scala,pines}.
E.g. one long standing puzzle has been the elucidation of the origin of the 
linear in temperature in-plane resistivity \cite{iye} 
and linear in 1/energy optical 
conductivity observed in the optimal doping regime and to a good extent
in the overdoped regime.
Vice-versa, the answer to this question should shed light on the character of
the carriers and, subsequently, on the superconducting transition.
Here we address these issues based on a {\em minimum} unconventional Fermi 
liquid model \cite{gk}. Our model comprises strong peaks in its density
of states (van-Hove singularities in 2 dimensions) near the chemical potential.
We show that it accounts in a natural, 
comprehensive and internally consistent manner
for several normal state characteristics. The introduction of an {\em
ansatz} for the susceptibility of the carriers further allows us both
to propose an explanation for the origin of the spin resonance peak
and to obtain
particularly high $d$-wave transition temperatures $T_c$.
Overall, our results make a strong case for a Fermi liquid approach
to the optimally doped and overdoped cuprates.

	We perform a combination of analytical and numerical many-body 
calculations in the context of our model.
The rest of this paper is organized as follows.	
In Section II we write our many-body approximation, which we used
in our numerical calculations. We emphasize both that
our treatment is relevant for overdoped and optimally doped 
cuprates (see Section VI for the underdoped regime) and that the results 
presented in Sections III-V depend only {\em quantitavely} and not 
qualitatively on the specific Hamiltonian and approximation thereof etc.
In Section III we discuss our analytical and numerical results
for the linear in max($T,\ep$) scattering rate of the carriers, 
in connection with the existence
of van Hove singularities (vHs) close to the chemical potential. We also show
that a conductivity linear in $(1/T,1/\ep)$ follows. In Section IV
we discuss both the fermionic origin of the energy dependence of the 
Millis-Monien-Pines susceptibility and our ansatz of eq. (\ref{ansatz})
for the susceptibility
of the carriers. We show that this ansatz may explain the appearance of the
so called spin resonance peak, seen in neutron scattering experiments,
in the normal state of the cuprates.
In Section V we discuss the superconducting transition 
in the frame of the ansatz. Finally Section VI contains a summary of our 
results. In the Appendix 
we examine the role of doping-induced disorder on the carrier 
susceptibility.

\vspace{1cm}
\centerline {\bf II. General framework}
\vspace{1cm}

	We assume that we deal with a Fermi liquid, albeit an
unconventional one, as will become apparent from our discussion 
of the scattering rate of the carriers below. 
We choose the 2-dimensional Hubbard Hamiltonian as a specific model
for our numerical calculations (c.f. the last paragraph of Section I
on this).

\be
H = \sum_{k,\sigma} \epsilon_k \; c_{k,\sigma}^{\dag} c_{k,\sigma}
+ \frac{U}{2 N^2} \sum_{k,k',q,\sigma} c_{k+q,\sigma}^{\dag} 
c_{k'-q,-\sigma}^{\dag} c_{k',-\sigma} c_{k,\sigma}  \;\;.
\ee
$c_{k,\sigma}^{\dag}$ is an electron creation operator and 
$\epsilon_k$ is the electronic tight-binding dispersion suggested by 
angle-resolved photoemission (ARPES) experiments - e.g. see \cite{stoj} - and 
LDA 
calculations \cite{ander}
\be
\epsilon_k = -2 t (\cos{k_x}+\cos{k_y})-4t'\cos{k_x}\cos{k_y}-
2t''(\cos{2k_x}+\cos{2k_y}) \;\; .
\ee
It is assumed here that the lattice constant 
is equal to unity and is the same along the two
crystal axis $a,b$ in the planes - hence $k_x,k_y \in [-\pi,\pi]$. 
$N\times N$ is the discretization of the Brillouin zone.

We consider the fluctuation-exchange diagrammatic approximation (FLEX)
of Bickers, Scalapino and White \cite{flex} for the Hamiltonian, which
consists in summing bubble and ladder diagrams. 
FLEX is a Baym-Kadanoff
conserving approximation \cite{bk},
meaning that there is a free energy
functional $\Phi[G]$ of the Green' function $G$, such that the self-energy
$\Sigma$ is given by the relation $\Sigma = \delta \Phi[G]/\delta G$.
We thus obtain a set of
self-consistent equations for $G(k,\ep_n)$ and $\Sigma(k,\ep_n)$ :
\be
G(k,\ep_n)^{-1} = G_o(k,\ep_n)^{-1} - \Sigma(k,\ep_n)  \;\;, 
\ee
\be
G_o(k,\ep_n) = \frac{1}{i \ep_n + \mu - \ep_k } \;\; ,
\ee
\be
\Sigma(k,\ep_n) = - \frac{T}{N^2}\sum_{q,\omega_m} V(q,\omega_m) G(k-q,\ep_n-
\omega_m) \; \; . \label{sigm}
\ee
The potential $V(q,\omega_m)$ is given by
\be
V(q,\omega_m)=V_{ex}(q,\omega_m)-V_H(q,\omega_m) \;\;, \label{dyn}
\ee
\be
V_{ex}(q,\omega_m) = \frac{ - U^2 \chi_o(q,\omega_m)}
{1 - U^2 \chi_o^2(q,\omega_m)} \;\; ,
\ee
\be
V_H(q,\omega_m) = \frac{U^3 \chi_o^2(q,\omega_m)}
{1 - U \chi_o(q,\omega_m)} \;\;, 
\ee
The susceptibility $\chi_o(q,\omega_m)$ is given by 
\be
\chi_o(q,\omega_m) = -(T/N^2) \sum_{\ep_n,k} G(k+q,\ep_n+\omega_m) G(k,\ep_n) \;
\; .
\label{epid}
\ee
$\mu$ is the chemical potential and the Matsubara frequencies
are $\ep_n=(2n+1)\pi T$ and $\omega_m=2m\pi T$ for fermions and bosons,
respectively.
We solve numerically this self-consistent set of equations,
working with a given number $M$ of Matsubara frequencies
and discretization of the Brillouin zone
($M=256-480$ and $N \geq 64$).

There have been a number of similar numerical calculations on the normal-phase 
and the superconducting transition of the cuprates
\cite{flex,mont,sglb,vilk,carb,andersen}, 
with FLEX 
being a particularly popular approach.

	All the convolution operations are done by using the Fast Fourier
Transform (FFT), in order to cut down calculation time. 
We use Pad\'e approximants \cite{pade} to analytically continue 
our results to the real frequency axis.

\vspace{1cm}

\centerline {\bf III. On the scattering rate of the cuprates}

\vspace{1cm}

	We have analytically obtained a scattering rate linear in
the maximum of the temperature $T$ or energy $\epsilon$,
for a Fermi liquid with strong density of states peaks - van-Hove
singularities (vHs) in 2-d - located at an energy $\epsilon_{vH}$
close to the chemical potential $\mu$.

	The derivation relies on the relation (\ref{sigm}) for the 
self-energy, which is valid quite generally in the frame of a BK
approximation, irrespectively of the specific Hamiltonian and approximation
thereof. See also the discussion following eq. (\ref{imv}).
	It can easily be shown \cite{agd} that $Im \Sigma(k,\ep)$ 
is given by the following formula at finite temperature :
\be
Im \Sigma^R(k,\ep) = \sum_{q,\om} Im G^R(q,\ep-\om) Im V^R(k-q,\om) \; 
\{\coth(\om/2T) \; + \; \tanh((\ep-\om)/2T) \}\;\;.
\ee
Taking 
\be
Im G^R(k,\ep) = -\pi \delta(s_{k,\ep}), \; \; s_{k,\ep} = \ep+\mu-\ep_k-Re
\Sigma(k,\ep) \;\; ,
\label{gde}
\ee
we obtain
\be
Im \Sigma^R(k,\ep) = -\pi \sum_q Im V^R(k-q,s_{q,\ep}) \;
\{ \coth(s_{q,\ep}/2T) \; + \; \tanh((\ep-s_{q,\ep})/2T) \}\;\; .
\ee
Setting $Im G^R(k,\ep)$ equal to a delta function is a reasonable
approximation for this purpose, in view of the typical sharp spike feature
of $Im G^R(k,\ep)$ shown in fig. 1 - also see figs. 2 and 3, all of which 
are representative of our numerical solution of eqs. (3)-(9).
Further, numerically $Im G^R(k,\ep)$ is 
{\em very small} compared to the band energy for small couplings,
and the difference of $Im \Sigma(k,\ep)$, as seen in our numerical
calculation, for small and large coupling constants is mostly {\em quantitative}
rather than qualitative. 

	We write
\be
Im V^R(q,x) = \sum_{n=0}^{\infty} \; \frac{V_q^{(2n+1)}(0) \; x^{2n+1}}{(2n+1)!}
\; ,
\ee
where $V_q^{(n)}(0)$ is the $n-$th derivative of $Im V^R(q,\om=0)$ with respect
to $\om$. This is true for an electronically mediated interaction,
with a polarization which is a regular function of $\om$ (see also eq. 
(\ref{imv}) below).
There are only odd powers of $\om$ in the series because the imaginary
part of the susceptibility is an odd function of energy - e.g. c.f. 
eq. (2.63) of Pines and Nozi\`eres \cite{pino}.
One possible exception to this is given by Gonzalez, Guinea and Vozmediano
\cite{gon}. The authors showed that for the {\em underdoped LSCO-type} 
Fermi surface (FS)
and for momenta $K$ connecting two inflection points of the FS, the imaginary
part of the susceptibility goes like $|\om|^{1/4}$ in 2 dimensions. However,
this fact will influence the final result for the scattering rate {\em only} for
a small range of momenta $k$ satisfying $k=K+q_o$ - c.f. eqs. (14) and (15),
and $q_o$ is given in the paragraph below. Moreover, we have already emphasized
that our picture is not valid in the underdoped regime.

	First we consider the low $T$ limit. 
The sum over $q$ is dominated by the van-Hove singularities at the points 
$q_o$. Assuming that $\ep_{vH} = \ep_{q_o} + <Re \Sigma(q_o,\ep)>$ 
[{\em this relation is misprinted in the journal version of the paper}]
is {\em close} to $\mu$, the
tanh has a vanishing contribution at the vicinity of 
$\ep_q + Re \Sigma(q,\ep) \sim \mu$
(note that for $\ep_q + Re \Sigma(q,\ep) < \mu$ 
and $\ep_q + Re \Sigma(q,\ep) > \mu + \ep$ the contributions
of tanh and coth annihilate each other in the low $T$ limit). Hence
\be
Im \Sigma^R(k,\ep) \simeq -\pi \sum_{q \sim q_o} \sum_{n=0}^{\infty} 
\frac{ V_{k-q}^{(2n+1)}(0) \; (s_{q,\ep})^{2n+1}}{(2n+1)!} \; \;,
\ee
For {\em sufficiently small} $V_{q}^{(n)}(0), \; \forall n>1$, we obtain
\be
a_k = \pi\sum_{q \sim q_o} V_{k-q}^{(1)}(0) \; \gg \; \pi\sum_{q \sim q_o} 
\sum_{n=1}^{\infty} \; \frac{ V_{k-q}^{(2n+1)}(0) \; (\ep + c)^{2n}}{(2n+1)!} 
\;\;,
\ee
where $c = \mu - \ep_{vH}$. This relation is valid for $\ep + c < \ep_c$,
where the latter is the characteristic energy beyond which the infinite sum
on the right becomes comparable to the $V_{q}^{(1)}$ term. 
Also, for energies beyond the bandwidth $W$ ($W=8t$ for the non-interacting
system), $Im \chi_o$, and hence 
$Im V$ (see below), decay to zero.
These considerations yield the two energy crossovers
\be
\ep_1 = |\mu - \ep_{vH}| \;\;,\;\;
\ep_2 = \min\{\ep_c + \ep_{vH} - \mu, \; W + \ep_{vH} - \mu\} \;,\;\; 
\ee
while the assumption above for $a_k$ leads to
\be
Im \Sigma^R(k,\ep) \simeq - a_k (\ep + c) \;\;, \;\;\ep_1 <  \ep  
< \ep_2 \;\;.
\label{taye}
\ee
For $\ep > \ep_2$ $Im \Sigma$ gradually decreases, due to the finite
bandwidth of the system.
Finally, we note that if the Fermi surface approaches a van-Hove singularity
at $q_o$,
$V_{k-q}^{(1)}(0)$ should become bigger, being proportional
to $1/(\vec \nabla \ep_{k_F} \vec k_F$) (as implied by the standard Fermi 
liquid
result for the imaginary part of the susceptibility \cite{pino} - see the 
discussion below on the susceptibility of the cuprates).

	We consider now the high temperature limit $T > (\mu-\ep_{vH})/4$
\cite{varel}. 
We see immediately that 
\be 
Im \Sigma^R(k,\ep) = - \pi \sum_q Im V^R(k-q,s_{q,\ep}) \{ 2 T/s_{q,\ep} +
O(s_{q,\ep}/2T) \} \;\; .
\ee
(Note that the term of order $T$ of this sum is
reminiscent of the left-hand side of the sum rule - 
c.f. Pines and Nozi\`eres\cite{pino} - 
$\lim_{q \rightarrow 0} \int_0^{\infty} d\om Im \chi_o(q,\om) |\varepsilon
(q,\om)|^2/\om = - N\pi/m c_s^2 $, with $N$ being the total particle number,
$c_s$ the speed of sound, $m$ the effective mass, and $\varepsilon(q,\om)$
the dielectric function.)
The sum is dominated by the van-Hove singularities at the points
$q_o$, thus yielding
\be
Im \Sigma^R(k,\ep) \simeq - 2 T \pi \sum_{q \sim q_o} V_{k-q}^{(1)}(0)  = 
- 2 a_k T \;\;. \label{tayt}
\ee
Here we made use of the condition above for $a_k$. 
In addition, it is straightforward to see from our analytic
treatment that $Im \Sigma^R \propto x^2$, $x=$max$\{T,\ep\}$, when
both $T,\ep \rightarrow 0$. In all respects we have a
genuine Fermi liquid.

{\em Note added:} ARPES expts. by Valla et al., Science {\bf 285}, 2110 (1999),
and preprint cond-mat/0003407, 
have very recently shown that in optimally doped Bi$_2$Sr$_2$CaCu$_2$O$_{8+y}$ 
the one-particle scattering rate is linear in max\{$T,\epsilon$\} 
over most of the Fermi surface, in support of our picture.

	A brief comment here. It has been known long ago - see e.g. \cite{sil}
- that the scattering rate becomes linear in $T$ for $T > \om_B/4$, with 
$\om_B$ being the characteristic boson frequency mediating the carrier 
interaction. Our treatment shows that $\om_B$ here is nothing else but
the {\em fermionic} energy $\mu-\ep_{vH}$.

	The prefactors in the r.h.s. of 
eqs. (\ref{taye}) and (\ref{tayt}) differ by a factor of 2. 
This is in agreement with experiments in YBa$_2$Cu$_3$O$_{7-\delta}$ and
Bi$_2$Sr$_2$CaCu$_2$O$_{8+y}$ \cite{tanner}, where the factor is found to 
be in the range 2.1 - 2.2 .
We note that the "marginal Fermi liquid" phenomenology of Varma, Littlewood,
Schmitt-Rink, Abrahams and Ruckenstein  \cite{mfl} gives a factor 
of $\pi$ instead.

	We emphasize that the $T$ and $\epsilon$ dependence of the result are 
{\em independent} 
of $k$ - {\bf thus leading necessarily to a linear in $T$ resistivity and 
a linear
in $1/\ep$ optical conductivity,} even with inclusion of vertex corrections
in the calculation. 
The reason for this being that $\{T,\ep\}$ are obtained
as overall prefactors, for the relevant $T$ and $\ep$ regimes, in such 
calculations.
E.g. the Kubo formula yields
$\sigma(\om)=(e^2/\om) \sum_{k,\ep}  v_k^2 \; G(k,\ep+\om) G(k,\ep) 
\;[1+S(k,\ep)] \; [f(\ep+\om) - f(\ep)]$, 
where $v_k$ is the group velocity, 
$S$ includes vertex corrections
from the Ward identity, and $f$ is the Fermi occupation factor. 
The main $T$ and $\om$ dependence in the integrand is in
the one-particle self-energy in $G$ and in $f$. Doing the $k$ sum,
we get the dominant contribution from the poles of $G$. Now, the 
one-particle scattering rate is linear in max($T,\ep$) {\em everywhere}
in the Brillouin zone, and this linear dependence appears in the 
denominator of $\sigma$.
Here we assumed that the vertex corrections do {\em not} have a strong 
temperature dependence over a substantial part of the Brillouin zone.
Indeed, Kontani, Kanki and Ueda \cite{kku} have recently shown numerically,
in the frame of the FLEX approximation, that 
vertex corrections are small for the resistivity, and do {\em not} 
change its $T$ dependence.

	Hlubina and Rice \cite{hlu} considered analytically a
model of interacting fermions with a vHs close to $\mu$. However,
they find a scattering rate similar to ours only close to the vH region,
and different otherwise. As a result, their resistivity goes like
$T^2 \; ln^2(1/T)$. In their 'hot' and 'cold' spots scenario, relying 
on strong scattering off antiferromagnetic fluctuations, they 
obtain an {\em average} scattering rate similar to ours, and numerically
a linear in $T$ resistivity (however, they seem to {\em assume} that 
the group velocity is finite along the whole Fermi surface - cf. between
eqs. (2.5) and (2.6) of \cite{hlu}). Similar results are also obtained
in the antiferromagnetic scenario of Pines and Stojkovic \cite{stp}.

	A note on phonons. As they form a - presumably small- part of the
effective potential $V$, they provide necessary momentum dissipation, yielding
a finite resistivity. However, the linear $T$ dependence of the latter 
is {\em not} specifically influenced by phonons in our model.

	Returning to the derivation of the scattering rate above, 
we observe that the overall behavior of $Im V$ closely 
follows $Im \chi_o(q,\om)$, as
\be
Im V(q,\om) = Im \chi_o(q,\om) \;\; |\varepsilon(q,\om)|^2 \;\;, \label{imv}
\ee
$Im \chi_o$
is odd in $\om$, while $|\varepsilon|^2$ is even.
Eq. (\ref{imv}) follows from any screened interaction
between the carriers. Hence the argument for the
linear in energy and temperature behavior of $\tau^{-1}(T,\ep)$
is equally generic. It relies essentially on a large coefficient for the 
linear in energy term of $Im V$ - i.e. of $Im \chi_o$ - and 
the presence of van-Hove singularities
{\em near} the Fermi surface. The result holds {\em regardless} of the 
dimensionality of the system. However, it is important that a significant 
part of the spectral weight be included in the strong peaks of the density 
of states lying close to $\mu$.

What is more, in our self-consistent numerical solution 
we observe that the energy $\ep_{vH}$ of the singularities is pushed 
by the interactions close to the chemical potential
- see fig. 4. This result is especially pronounced when we use the ansatz 
for the susceptibility of the carriers of eq. (\ref{ansatz}) below.
Then we find typically for $n \sim 0.87 - 0.95$ and for a broad range
of $t',t'',U$ 
\be
\mu - \ep_{vH} \leq t/20 \;\;.
\ee
The shape of the 
self-energy $\Sigma(k,\ep)$ of the interacting system
is responsible for the modification of the density of states $N(\ep)$, 
through the relation $N(\ep)=-Tr \; Im G(k,\ep)/\pi$.
A trend for the transfer of the spectral weight is indicated by the fact that 
$Im \Sigma(k,\ep)$ has a peak below $\mu$ and a dip above it.
The numerical result concerning the approachment between $\ep_{vH}$ and $\mu$
has been known for some years.
Si and Levin \cite{sil} and Newns, Pattnaik and Tsuei \cite{newns} 
observed the pinning of the vHs close to $\mu$
by using a $U \rightarrow \infty$ mean field slave boson approximation of 
a model with Cu 3d and O 2p orbitals.
Recently, Gonzalez, Guinea and Vozmediano \cite{ggv} were able to obtain 
analytically the essential part of 
the approachment between the vHs and $\mu$ with a first order 
renormalization group treatment in the context of the Hubbard model.
A review of related work in the frame of the so-called van-Hove 
scenario has been given by Markiewicz \cite{mark}. 
This pinning of the vHs close to $\mu$ seems to be a plausible explanation
for the common characteristic of a good many cuprates whose van-Hove
singularities are located between 10-30 $meV$ {\em below} the Fermi surface
\cite{bednorz} (see also the next section).

	It is interesting that the electron doped 
Nd$_{2-x}$Ce$_x$CuO$_{4+\delta}$
which has a van-Hove singularity much below the Fermi surface, i.e. at 
approximately $\mu$-350 meV, as shown by ARPES\cite{stanford},
has a usual Fermi liquid $\tau^{-1}(T) = const. \; T^2 ln(T)$ \cite{tsuei}.
This lends support to the picture described above.
Along the same line, the resistivity of Tl$_2$Ba$_2$CuO$_{6+\delta}$ (Tl-2201)
switches over from linear to quadratic with increasing doping from the optimal
to the overdoped regime \cite{kubo}, which we suspect to be an indication
of the vHs moving well away from $\mu$.

        Finally, in further support of the relevance of the vHs in the
transport properties of the cuprates, Newns et. al \cite{newns2} and
McIntosh and Kaiser \cite{kaiser} have shown that the thermal conductivity
of the cuprates can be well accounted for if the vHs are located very close
to the Fermi surface, as discussed above.

	The numerical solution of the many-body system always
corroborates our analytical result for the self energy at finite temperature.
$Im \Sigma(k,\epsilon)$ turns out
to be essentially linear in energy in the interval
$\epsilon_1<\epsilon<\epsilon_2$.
A linear dependence of $Im \Sigma(k,\epsilon)$ as a function of either $T$
or $\epsilon$ was also obtained in the numerical work of Beere and Annett
\cite{beere}, Kontani et al. \cite{kku} and Si and Levin \cite{sil}.
Note that in figs. 2 and 3 we show the self-energy for the set of the system
parameters which yields the highest transition temperature $T_c$, if use
of the ansatz of eq. (\ref{ansatz}) is made. The linearity of 
$Im \Sigma(k,\epsilon)$ with $\ep$ is even {\em more} pronounced for 
other combinations of $t',t''$ and $n$.
$Im \Sigma(k,\epsilon)$ has always the correct parabolic Fermi liquid bevahior
for $\epsilon \rightarrow 0$.
Furthermore,
the energy interval of linear behavior expands as the energy $\epsilon_{vH}$
of the (extended)
van-Hove singularities at the (vicinity of the) points
$q_o =(\pm \pi,0), (0,\pm \pi)$ approaches $\mu$.
                               
	Another feature of the density of states as seen in our treatment
- c.f. fig. 4 - is the following. The non-interacting density of states has 
two minor 
peaks at the bottom and top of the spectrum respectively (in fig. 4 the top
one is a vHs). As the 
strength of the interaction increases, these two peaks are washed out, as
a result of the self-energy which becomes substantial in magnitude for 
energies away from $\mu$ - c.f. figs. 2 and 3.

At the moment it is not clear whether the present mechanism of the linear
scattering rate can explain the experimentally observed $T^3$ dependence of 
the Hall resistivity of the cuprates. A
way to explain it has been found by Stojkovic and Pines \cite{stp}, 
using an electron interaction peaked at $Q=(\pm\pi,\pm\pi)$. 
Their argument can be slightly modified, so that it works for our form
of the electron potential $V$ - given by eq. (\ref{dyn}) - but with a 
modified effective
$\chi_o$ peaked at Q, as we propose in the next section 
- c.f. eq. (\ref{ansatz}) and below.
Kontani et al. \cite{kku} have shown that vertex corrections in the frame of 
FLEX have a drastic influence on the $T$ dependence of the Hall resistivity,
in marked contrast to the case of the longitudinal resistivity.

\vspace{1cm}

\centerline{\bf IV. On the susceptibility of the cuprates}

\vspace{1cm}

{\em \bf The low energy dependence of the susceptibility of the cuprates.}
The  Millis-Monien-Pines susceptibility\cite{mmp,pines}
\be
\chi_{MMP}(q,\om) = \frac{X_1 \; \xi^2}{1+\xi^2 (q-Q)^2 - i\om/\om_{SF}} \;\;,
\ee
has been used to fit the {\em low} energy part of the 
susceptibility of the 
cuprates in both NMR rate and inelastic neutron scattering (INS) experiments.
Here $Q=(\pm\pi,\pm\pi)$.
The short range antiferromagnetic (AF) order, a remnant of the parent 
antiferromagnetic materials, with correlation length $\xi$, 
is responsible for the peak of the susceptibility for $q$ near $Q$.
Typically $\xi$ is of the order of the lattice constant ($\xi$ decreases
as the doping increases, and e.g. $\xi \simeq 2$ for optimally doped 
YBa$_2$Cu$_3$O$_{7-\delta}$), while $\om_{SF} \approx 10-40 meV$.

	The origin of the small magnitude of $\om_{SF}$ has remained 
elusive thus far.
E.g. Sachdev, Chubukov and Sokol have interpreted it as a damped spin
wave mode \cite{css}. Spin waves are clearly observable in underdoped cuprates.
However to date there is no experimental proof that they are strong
enough in the normal phase of the optimally doped and overdoped regimes. 
The proximity of the system to an antiferromagnetic instability, 
i.e. $\bar V_Q \; \chi(Q,\om) \simeq 1$, where $\bar V_q$ and $\chi(q,\om)$ 
are some
appropriate coupling and susceptibility respectively, can in principle
explain the small magnitude of $\om_{SF}$, as pointed out by Millis, Monien 
and Pines \cite{mmp}.

	Here we propose that an alternative explanation - which may coexist 
with the latter - is the following fermionic origin for $\om_{SF}$.

First however, let us present a proposal for the susceptibility which 
has been put forward
by Onufrieva and Rossat-Mignod\cite{onu}. This starts by viewing the CuO$_2$
planes as a lattice of plaquettes centered on the copper site with four 
nearest neighbour oxygen sites.
A Hamiltonian ${\cal H}$, reminiscent of but more comprehensive than
the one of the $t-J$ model, was introduced in terms of the Hubbard 
operators.
In this formulation, the itinerant carriers which propagate via Cu spin flips
are clearly {\em separate} objects from the localized Cu spins with short range
AF order.
A diagrammatic approach was developed in the frame of ${\cal H}$,
leading to the following RPA-type {\em total} susceptibility
\be
\chi_t(q,\om) = \frac{\chi_{AF}(q,\om) + \chi_F(q,\om)}{1 + 
J_q (\chi_{AF}(q,\om) + \chi_F(q,\om))} \;\;. \label{xon}
\ee
$J_q$ is the effective Cu spin exchange interaction, $\chi_F$ is a purely
fermionic susceptibility and $\chi_{AF}(q,\om)$ is 
due
to the localized spins. $\chi_t(q,\om)$ encompasses in an appealing way
the idea of the
entangled carrier-spin dynamics in the cuprates.
Furthermore, this approach is able to account to a good extent for the 
variation of the total susceptibility as a function of doping and temperature,
as measured by INS.

	Now, we use the result for $\chi_{t}$ above with
\be
\chi_{AF}(q,\om) = \frac{\chi_1 \; \xi^2 }{1+\xi^2 (q-Q)^2 - f(\om)} \;\;, \;\;
\chi_F(q,\om) = \chi_{Fo} (1+i \om/\om_o + O(\om^2)) \;\; , 
\;q \rightarrow Q \;\;.
\ee
Let us suppose that $f(\om)=i \om/\om_S$. 
If $\om_S \gg \om_o$ {\em and} $J_Q \xi^2 \chi_1 < 1$, 
and taking $\chi_F(q,\om) \equiv \chi_o(q,\om)$ (as given by eq. (\ref{epid})),
we essentially recover $\chi_{MMP}(q,\om)$ - which is
itself an {\em approximate} form of the true susceptibility - with 
\be
\om_{SF} \rightarrow \bar \om(q) = \frac{\om_o \; \om_S}{\om_o + 
\om_S J_q \chi_o^2 \; (1+\xi^2 (q-Q)^2)\;/\;(\xi^2 \chi_1)} \;\; . 
\ee

From the numerical solution
of our system, we easily obtain values of $\om_o$ comparable to 
the experimentally relevant ones, when the van-Hove energy $\ep_{vH}$ is near 
$\mu$, with $\om_o$ scaling quickly towards zero as $\mu - \ep_{vH} 
\rightarrow 0$.
Hence $\om_o$ can be interpreted as $\om_o(\vec q_F)=\vec \nabla \ep_{q_F} 
\vec q_F$ - c.f. the non-interacting Fermi liquid result $\om_o(q)=v_F q$
\cite{pino}.
The small difference $\mu - \ep_{vH}$ is observed in a good number 
of cuprates.
E.g. in ref. \cite{bednorz} there is a compilation of several cuprates,
the van-Hove singularities of which are located between 10 - 30 $meV$ 
below the Fermi level (c.f. the discussion in the previous section).
Also, Blumberg, Stojkovic and Klein (BSK) \cite{stoj} suggested
that this characteristic may be true irrespective of the doping,
as long as the latter is appropriate for superconductivity.
This is based on ARPES experiments on the bilayer YBa$_2$Cu$_3$O$_{7-\delta}$. 
ARPES remains the best diagnostic probe
for the Fermi surface of the cuprates. 
Yet it has not proved possible to perform measurements
on many other compounds, especially the monolayers 
such as Tl$_2$Ba$_2$CuO$_{6+\delta}$ etc.
The point here is the following. By fitting the ARPES data BSK
show that one of the two effective bands - the anti-bonding one - formed 
by hybridization of the two layers by interlayer coupling has a chemical
potential only some 20 - 50 $meV$ above the van-Hove singularity at
$(0,\pi)$, irrespective of the doping regime. It is then clear that these
carriers, with a large density of states, give rise to a {\em small} $\om_o$ 
as discussed above. Hence it is very 
interesting to know 
how universal this band-structure characteristic of the cuprates is,
as it may explain naturally the magnitude of $\bar \om$.
Furthermore, it would be interesting to determine experimentally,
e.g. by 
INS, the value of $\bar \om$ for Nd$_{2-x}$Ce$_x$CuO$_{4+\delta}$.
In that case, $\om_o$ should be enhanced as a result of the van-Hove 
singularities being far away from the Fermi surface.

In the Appendix we discuss the (non)influence of weak disorder on the value of
$\om_o$.

{\em \bf The antiferromagnetic ansatz for the carrier susceptibility.}
We thereby propose that the {\em effective} non-interacting
susceptibility (i.e. without interaction lines connecting the particle-hole 
lines) of the carriers is given by the following {\bf ansatz}
\be
\chi_o(q,\om) \rightarrow \chi_o^{eff}(q,\om) = \chi_o(q,\om) + 
a \; \chi_{AF}(q,\om) \;\;.  \label{ansatz}
\ee
$\chi_o$ is given by eq. (9) above, 
$\chi_{AF}$ is the antiferromagnetic susceptibility of the localized
Cu spins and $0 < a < 1$ is a dimensionless weighting factor, which in
principle depends on doping, band structure, temperature (presumably a
decreasing function of the latter) etc. 
The intuitive idea is that the carriers
should perceive the ordered antiferromagnetic background of the CuO$_2$ planes, 
even in the absence
of phase separation \cite{ek,zaa,cn}.
In this way, the fermionic susceptibility acquires an antiferromagnetic
enhancement, which may then influence the effective carrier potential
and pairing, through an electronically
mediated interaction - see also the next section. Furthermore,
$\chi_o^{eff}$ can explain the temperature dependence of the Hall resistivity
of the cuprates, as we mentioned at the end of the previous section
on the scattering rate.
Finally, we note that with this effective $\chi_o^{eff}$ our many-body 
approximation remains {\em conserving}, since the relation $\Sigma = \delta F/
\delta G$ between the self-energy, the free energy and the Green's function
is still valid, as $\chi_{AF}$ does {\em not} depend on $G$.
We note that this is consistent with the work of Onufrieva and 
Rossat-Mignod \cite{onu} mentioned
above. The carriers and the localized spins form two distinct, albeit 
interrelated, systems. The additive form of the ansatz is also compatible 
in spirit with $\chi_t(q,\om)$ of eq. (\ref{xon}).

We mention here the 
alternative treatment of the spin and charge susceptibilities of
Imada, Fujimori and Tokura \cite{imad}.

Taking the ansatz of eq. (\ref{ansatz}) at face value in the context of 
our many body 
scheme means that both the charge and spin susceptibility of the carriers 
acquire an AF enhancement. Currently we cannot prove this ansatz. 
We emphasize that our ansatz can be taken to apply {\em only} 
for the spin
susceptibility of the carriers. In that case, in the frame of our 
Baym-Kadanoff scheme, only $\chi_o(q,\om)$ entering the ladder diagrams
with opposite particle-hole spins 
would be replaced by $\chi_o^{eff}$.
Quantitatively, the differerence between this case and the case in which both
charge and spin susceptibilities are enhanced is small (for relevant values
of the AF enhancement) - c.f. the discussion
on the critical temperature $T_c$ in the next section.

In our numerical implementation, we consider two similar forms 
for the AF susceptibility, namely (A)
\bea
\chi_{AF}^A(q,\om_m) = X_o \; \sum_{i=1}^4 \Gamma_i^{-1}\;
\theta(\omega_c - |\omega_m|) , \; \label{xx1}
\eea
with $\Gamma_i=\xi^{-2} + (q-Q_i)^2$, $\om_c$ being a cut-off, and (B)
\be
\chi_{AF}^B(q,\om_m) = X_o \; D \; \sum_{i=1}^4 \frac{ \om_m - (2 \om_m /\pi)
 \arctan(\om_m/D) - \Gamma_i D + (2 \Gamma_i D /\pi) \arctan(\Gamma_i) }
{\om_m^2 - (\Gamma_i D)^2} \;,  \label{ansb}
\ee
with $Q_i=(\pm\pi,\pm\pi)$ and $D$ being a cut-off frequency,
above which $Im \chi_{AF}(q,\om)=0$.
Form (B) has appeared in \cite{millis}. Here the characteristic spin wave
frequency obeys $\omega_S \propto \xi^{-z}$, and the $z=2$ scaling regime
has been assumed, in agreement with the analysis of Sokol and Pines
for the optimally doped and overdoped regime of YBa$_2$Cu$_3$O$_{7-\delta}$
etc. \cite{sokol}

{\em \bf The spin resonance peak of the cuprates.}
INS experiments in YBa$_2$Cu$_3$O$_{7-\delta}$ - see e.g. \cite{fong1} and 
therein - and Bi$_2$Sr$_2$CaCu$_2$O$_{8+y}$ \cite{fong2} have revealed 
the existence of a strong peak in the spin triplet channel, centered
at $Q=(\pm\pi,\pm\pi)$ and at a characteristic resonance energy
$\omega_R\simeq 30-40$ meV. Although this peak is usually seen exclusively
in the superconducting state, it has been observed up to temperatures
$\sim 250 ^o$K in YBa$_2$(Cu$_{0.995}$Zn$_{0.005})_3$O$_7$ \cite{fong1}
for $\omega_R\sim 40$ meV.
Interestingly, this fact cannot be accounted for by most of the theoretical
models so far available - see e.g. \cite{fong1,fong2} for refs. - as these
models require the onset of superconductivity. An exception is the 
model of Bulut \cite{bulut}, which, however, differs drastically from ours.

The use of the ansatz of eq. (\ref{ansatz}) in the spin channel only 
may account in a natural way for the appearance of the resonance peak
in the normal state through 
a bilayer effect. Both YBa$_2$Cu$_3$O$_{7-\delta}$ and 
Bi$_2$Sr$_2$CaCu$_2$O$_{8+y}$ are bilayer materials.
Here one can define bonding and antibonding bands \cite{andersen}
$\epsilon_{\pm k}=\epsilon_k\pm t_\perp(k)$, $t_\perp(k)$ 
being the $k$-dependent interlayer
hopping element. Further, one defines the susceptibilities
$\chi_{o \pm}(q,\om)= Tr[ G_+(+q,+\om) G_\pm +G_-(+q,+\om) G_\mp ]$.
When $\epsilon_{vH}-\mu$ is {\em small}, $\chi_{o-}(q,\om)$ has a 
narrow peak at $q=Q$ and 
$\om\simeq\Delta$, where $\Delta$ is the bilayer splitting at 
$q\sim q_o = (\pi,0),(0,\pi)$, i.e. the
van Hove neighbourhood. Liechtenstein et al. \cite{andersen} have shown in
the frame of the FLEX approximation that $\Delta$ is drastically reduced 
for a finite interaction, compared to the non-interacting value
$\Delta_o=2 t_\perp(q=q_o)$, such that it becomes comparable to the
experimental $\omega_R$. Our ansatz can be taken to apply to $\chi_{o\pm}$,
yielding  $\chi_{o \pm} \rightarrow \chi_{o \pm}^{eff}=\chi_{o \pm} 
+ a\; \chi_{AF}$.
As a result $\chi_-(q,\om)=\chi_{o-}^{eff}(q,\om)/(1-U \chi_{o-}^{eff}(q,\om))$ 
is strongly peaked at $Q$ for an energy $\om_{cR}\sim\Delta$, and may 
account for the experimental observations.
Of course we require $\chi_{o-} > \chi_{o+}$ for this to work, which 
is valid for a range of the parameters $t',t'',n$.

One can ask the question: why does the resonance peak not appear in the 
normal state in general? As demonstrated here, the amplitude of the AF
enhancement needs to be sufficiently large for the peak to be visible.
Zn is known to enhance AF fluctuations
in the CuO$_2$ planes - e.g. \cite{casba}, which if interpreted as yielding 
a larger ($a \; \chi_{AF}$) contribution
in $\chi_o^{eff}$, is in agreement with the results above.
On the other hand, as we discuss in the next section on the superconducting
transition, too strong a factor $(a \; \chi_{AF})$ reduces 
$T_c$ - c.f. $T_c=93^o$K
for the pure YBa$_2$Cu$_3$O$_7$ versus $T_c=87^o$K for the Zn doped material
mentioned above (also see the last paragraph of section V).

\vspace{1cm}

\centerline{\bf V. On the superconducting transition of the cuprates}

\vspace{1cm}

	With the solution of the normal system at hand, we solve
the gap (Eliashberg) equation\cite{loh} 
\be
\Delta(k,\ep_n)=-\frac{T}{N^2}\sum_{k',\ep_n'} V_{p}(k-k',\ep_n-\ep_n') 
G(k',\ep_n')G(-k',-\ep_n')\Delta(k',\ep_n') \;\; . \label{gap}
\ee
This form of the equation is valid close to the transition temperature 
$T_c$ only.  
The pairing potential $V_{p}$ is given by
\be
V_{p}(q,\omega_m) = V_{x}(q,\omega_m)  + V_h(q,\omega_m)  \;\;,  
\ee
\be
V_{x}(q,\omega_m) = \frac{U}{1 - U^2 \chi_o^2(q,\omega_m)} \;\;,
\ee
\be
V_h(q,\omega_m) = \frac{U^2 \chi_o(q,\omega_m)}
{1 - U \chi_o(q,\omega_m)} \;\;.
\ee
We have assumed that the gap is an even function of both its
momentum and energy arguments in order to write $V_{p}(q,\om_m)$ 
in this form. We also assumed that the gap is spin singlet, as
Knight shift experiments have shown \cite{scala,agl}. The superconducting
state of the cuprates is probably a generalized BCS state - see 
e.g. \cite{nayak} - 
with the transition being due to a momentum anisotropic potential,
as envisaged by Kohn and Luttinger \cite{kohn}.

	The symmetry of the gap is determined by the 
exact shape and sign of $V_p$.
Based on symmetry grounds \cite{scala} as
well as on experimental evidence, we expect a $d_{x^2-y^2}$ or
a $s$ wave gap - but see also \cite{shf}. 
(In principle, higher order even angular momentum harmonics are also possible.)
The highest $T_c$'s correspond to a $d_{x^2-y^2}$ gap, and are 
obtained by including an
antiferromagnetically enhanced susceptibility in the calculation, following our
{\em ansatz} of eq. (\ref{ansatz}).
In passing, let us note that in general the proximity of $\ep_{vH}$ to
$\mu$ plays a less significant role than the ansatz in raising $T_c$ - but
c.f. below.
Nevertheless we also obtain a $s$-wave gap under the same conditions,
but with a much lower $T_c'$ \cite{tran}. 
This is consistent with the experimental situation. Most of the cuprates
appear to have a $d_{x^2-y^2}$ gap at $T_c$ - e.g. c.f. \cite{agl}. 
Experiments point to the opening of a {\em secondary} order parameter
gap at $T_c' \ll T_c$ \cite{covi,kouz,movs}.

        We note here that the van-Hove singularities at $q_o$ tend to
suppress a
$s$-wave gap if the pairing potential $V_p>$0 ($V_p$ defined above
is negative for sufficiently large and negative $U$ and/or an appropriate
phonon coupling,
and positive otherwise).
The (likely) $s$-wave gap of Nd$_{2-x}$Ce$_x$CuO$_{4+\delta}$
could originate from e.g. 
the fact that the van-Hove singularities are 350$meV$
below the Fermi surface, and hence ineffective here, 
or from $V_p<$0 for some relevant parts of the phase space
in this material, 
or possibly from both facts.
Yet another possibility which may coexist with the above 
is that $V_p$ is strongly peaked close to zero momentum owing to
the band structure of this material. All these factors can
result in the $s$-wave $T_c$ being higher than the $d$-wave $T_c$, and hence in
the dominance of the former channel over the latter.

	We obtain the 
following (near) optimum set of parameters for the $d$-wave $T_c$.
We take $t=250 meV$ and $\xi=1$ (increasing the latter leads to a 
reduction of $T_c$, see below).
For form (A) - eq. (\ref{xx1}) - we obtain $T_c \simeq 125^o K$, 
for $a X_o=0.2eV^{-1}$,
$\om_c=4t$, $t'=-0.11t$, $t''=0.5t$, $U=1.27eV$ and $n=0.88$.
For form (B) - eq. (\ref{ansb}) - we obtain $T_c \simeq 105^o K$, 
for $a X_o=0.5eV^{-1}$,
$D=32t$, $t'=-0.11t$, $t''=0.25t$, $U=1.47eV$ and $n=0.91$.
The variation of $T_c$ in $^oK$ with $\xi$ is as follows for this last set
of parameters : (94, 1.5), (91, 2), (90, 3), (88, 5).

Monthoux and Pines obtained similarly high transition temperatures
\cite{mont} with their approach, which uses
${\cal V}(q,\om)=g^2 \chi_{MMP}(q,\om)$ as the effective carrier-carrier
potential. Their approach has no room for the Hubbard $U$ and the subsequent
screened carrier-carrier interaction though. Overall, our approach yields
significantly enhanced $T_c$'s compared to the standard FLEX-type
treatments \cite{fls}.

	We obtain an optimum value of $n$ for the following reason.
$\chi_o$ - and hence $\chi_o^{eff}$ - is a decreasing function of $n$.
We allow the coupling to increase up until 
$b = U \chi_o^{eff}(Q,\om=0)$ saturates to a value close to and below unity.
Then $V$ and $V_p$ (given by eqs. (6) and (32) respectively) are 
increasing with $n$. And so does the characteristic
scattering rate entering $G$. Thus the optimum value of $n$ for the
highest $T_c$ corresponds to the overall highest kernel $V_p \; |G|^2$
in the gap equation (\ref{gap}).

	Further, we have done an {\em extensive} search of the parameter space 
to locate the optimum parameter set for the highest $T_c$. From our data
it appears that the variation of $T_c$ as a function of the system parameters
is {\em smooth} and that the optimum parameter set above is a {\em global} 
optimum \cite{phon}.

	Assuming that only the spin susceptibility acquires an AF 
enhancement according to our ansatz, yields a $T_c$ which is lower
by 
9\% for $\xi=1$, but only lower by 2\% for $\xi=3$,
if we make use of $\chi_{AF}^B$, with the optimum parameters above.
The variation of $T_c$ with $\xi$ is as follows here :
(96,1), (92,1.5), (90,2), (89,3), (88,5).

There is an {\em optimum} value of $a$ for the maximum attainable $T_c$. 
This is again due to the form of the pairing potential $V_p$ above : 
the AF instability condition allows $U \chi_o^{eff} < 1$ only.
Now, for a given value of the latter product, the highest $V_p$ - which 
in principle yields the highest $T_c$ as well (c.f. above) - 
will correspond 
to the highest possible $U$. This in turn corresponds to a smaller 
$\chi_o^{eff}$, and hence to a {\em small} but finite optimum $a$.
In the next paragraph we discuss relevant experimental evidence.

	Zheng, Kitaoka, Ishida and Asayama \cite{zheng} made 
a very interesting empirical
observation. Namely, among the hole doped cuprates, the highest $T_c$ 's
correspond to a combination of {\em both} optimum total carrier concentration
$n_{x^2-y^2}+2 n_{p_{\sigma}}$ in the planes ($n_{x^2-y^2}$ being the 
concentration
of holes of Cu-3d orbital character and $n_{p_{\sigma}}$ of O-2p orbital 
character)
as well as of a reduced (probably minimum) imaginary susceptibility - 
as deduced from NMR experiments.
In fact, Zheng et al. noticed that the highest $T_c$ 's are obtained
for a reduced ratio $n_{x^2-y^2}/2n_{p_{\sigma}} \sim 1$, and, moreover, that
such a trend is correlated with a reduced relaxation rate $(1/T_1)_{Cu} \propto
T \; \lim_{\omega \rightarrow 0} \sum_{q} A_q^2 \; Im \chi(q,\omega)/\omega $,
in properly normalized units (here $A_q$ is the hyperfine coupling).
This last fact points to a reduced $Im \chi(q,\omega)$ etc.
These conclusions are in accordance with our picture, which yields both
special values of the filling factor $n$ - for this also c.f. e.g. 
\cite{flex,sglb} - as well as special {\em small} 
values of the product $(a \; \chi_{AF})$ as a prerequisite 
for the highest attainable $T_c$ 's (also c.f. the last paragraph of section IV).

\vspace{1cm}

\centerline{\bf VI. Summary}

\vspace{1cm}

        To summarize, we present a single plane Fermi liquid model
which for the normal state can explain the salient transport properties,
the low energy
dependence of $\chi_{MMP}$, and their relation to the existence
of van-Hove singularities close to the Fermi surface. 
E.g. we obtain analytically a scattering rate linear in max($T,\ep$), within
appropriate $T$ and $\ep$ bounds, for {\em all} momenta in the Brillouin
zone. This result yields directly a linear in $T$ resistivity and linear in 
1/$\ep$ optical conductivity.
The introduction of an ansatz for the susceptibility of the carriers allows 
for an understanding of both the appearance of the spin resonance peak in the
normal state and the temperature behaviour of the Hall conductivity.
Further, by using the ansatz we obtain significantly enhanced $d_{x^2-y^2}$ 
wave transition temperatures $T_c$. 
Attention is paid
to Nd$_{2-x}$Ce$_x$CuO$_{4+\delta}$, the properties of which,
despite appearances,
we believe to be fully consistent with those of the majority of cuprates.
          
        In brief, let us discuss the possible connection to the physics
of the underdoped cuprates. Strong experimental evidence suggests that they are 
in a phase separated
regime, with AF domains of spins separated by stripes of holes \cite{ek,zaa,cn}.
One can envisage that with doping increasing towards the optimal regime,
the stripes melt into an effective Fermi liquid, and the physics described
here is recovered.
Further models on the underdoped cuprates can be found in \cite{timu}.
               
        The author has enjoyed discussions with
Yia-Chung Chang, Gordon Baym, Joseph Betouras,
Girsh Blumberg, Antonio Castro Neto, Lance Cooper, 
Sasha Liechtenstein, Peter Littlewood, 
J\"{o}rg Schmalian, Raivo Stern, Qimiao Si and
Branko Stojkovic.
This work was supported by the Research Board of
the University of Illinois, the Office of Naval Research under
N00014-90-J-1267 and NSF under DMR-89-20538.

\vspace{1cm}
          
\centerline{\bf Appendix - Disorder effects in the susceptibility
of the carriers} 

\vspace{.5cm}
          
	Returning to the origin of a small $\om_o$, which was discussed
in section IV,
another option would in principle be the disorder inherent in the cuprates.
The dopants are randomly positioned in the crystal structure, thereby 
creating an effective disorder potential for the carriers in the planes.
We calculated the effect of disorder by considering the dopants as
isotropic point scatterers, with a density $n_i$=1\% and typical scattering
strength $V_s=0-1$ eV (i.e. $\le 8t$). These parameters give a residual 
impurity scattering rate less than 2 $meV$, consistent with experiments on 
the cuprates.
For the calculation of the
susceptibility we used the diffuson \cite{lera}.
We only used a band
structure with $t''=0$, and only took into account non-magnetic disorder. 
The non-interacting Green's function now becomes
\be
G_o'(k,\ep_n) = \frac{1}{i \ep_n + \mu - \ep_k + \sigma_i(\ep_n)} \;\; ,
\ee
with $\sigma_i(\ep_n) = n_i V_s/(1-(V_s/N^2) \sum_k G_o(k,\ep_n))$.
The susceptibility is given by 
\be
\chi_o(q,\omega_m) = - T \sum_{\ep_n} P(q,\omega_m;\ep_n) \Big\{
\frac{ \theta(-\ep_n(\ep_n+\omega_m))}{1- n_i V_s^2 P(q,\omega_m;\ep_n) }
\; + \; \theta(\ep_n(\ep_n+\omega_m)) \Big\} \;\; , 
\ee
and $P(q,\omega_m;\ep_n) = (1/N^2) \sum_k G(k+q,\ep_n+\omega_m) G(k,\ep_n)$.
However we found {\em no} evidence, for the parameters
above, of $\om_o$ being influenced by disorder.

\vspace{1cm}
$^*$ Current address: 18 Giampoudi St., Iraklio, Crete 71201, Greece.
E-mail: kast@iesl.forth.gr .

\vspace{.5cm}

\centerline{ FIGURE CAPTIONS}

\vspace{.5cm}

 Figure 1. Full Green's function $G^R(k_F,\ep)$ for (a)
$k_F=k_F^{boundary}$
along $(\pi,0) \rightarrow (\pi,\pi)$ and (b)
$k_F=k_F^{diagonal}$
along $(0,0) \rightarrow (\pi,\pi)$, for $t=250meV$, $t'=-0.11t$,
$t''=0.25t$, $U=1.5eV$, $n=0.91$, at $T=105 ^oK$. 

\vspace{.5cm}

 Figure 2. Self-energy $\Sigma(k_F,\ep)$ for
$k_F=k_F^{boundary}$
along $(\pi,0) \rightarrow (\pi,\pi)$ (continuous line) and
$k_F=k_F^{diagonal}$
along $(0,0) \rightarrow (\pi,\pi)$ (dashed line), for the same
parameters as in fig. 1.
(a) depicts Im$\Sigma^R(k_F,\ep)$ and (b) depicts Re$\Sigma(k_F,\ep)$.
A (quasi)linear energy dependence of Im$\Sigma^R(k_F,\ep)$ can be seen 
for energies below 0.5$eV$ - also c.f. text. 

\vspace{.5cm}

{Figure 3. Self-energy $\Sigma(k_F,\ep)$ for
$k_F=k_F^{boundary}$
along $(\pi,0) \rightarrow (\pi,\pi)$ (continuous line) and
$k_F=k_F^{diagonal}$
along $(0,0) \rightarrow (\pi,\pi)$ (dashed line), for $t=250meV$, $t'=-0.11t$,
$t''=0.25t$, $U=1.47eV$, $n=0.91$, at $T=105 ^oK$. 
The carrier susceptibility has an antiferromagnetic enhancement
according to the ansatz of eq. (\ref{ansatz}) here (see section IV), with $\xi=1$,
$aX_o=0.5eV^{-1}$ and $D=32t$. This is the optimum $T_c$ case when using
form (B) - eq. (\ref{ansb}) - of our ansatz.}
(a) depicts Im$\Sigma^R(k_F,\ep)$ and (b) depicts Re$\Sigma(k_F,\ep)$.
A (quasi)linear energy dependence of Im$\Sigma^R(k_F,\ep)$ can be seen 
for energies below 0.5$eV$. 

\vspace{.5cm}

{Figure 4. Evolution of the density of states. Dashed line : non-interacting
system, with parameters as in fig. 3. Continuous line : 
same system with $U=0.8eV$.
Notice the transfer of the central van-Hove peak towards the chemical
potential, the disappearance of the two satellite peaks - see text,
 and the broadening of the total spectral width.
}


\begin{figure}
\epsfbox{fi1a}
\end{figure}

\begin{figure}
\epsfbox{fi1b}
\end{figure}

\begin{figure}
\epsfbox{fi2a}
\end{figure}

\begin{figure}
\epsfbox{fi2b}
\end{figure}

\begin{figure}
\epsfbox{fi3a}
\end{figure}

\begin{figure}
\epsfbox{fi3b}
\end{figure}

\begin{figure}
\epsfbox{fi4}
\end{figure}


\begin{references}
\bibitem{phon}
We note that the basic effect of considering moderate values of an isotropic
phonon interaction 
is to allow for greater 
values of $U$ before 
the AF instability condition $b=1$ (c.f. above) sets-in.
Hence a finite phonon interaction 
allows for a bigger on-site/paramagnon coupling.
\end{references}
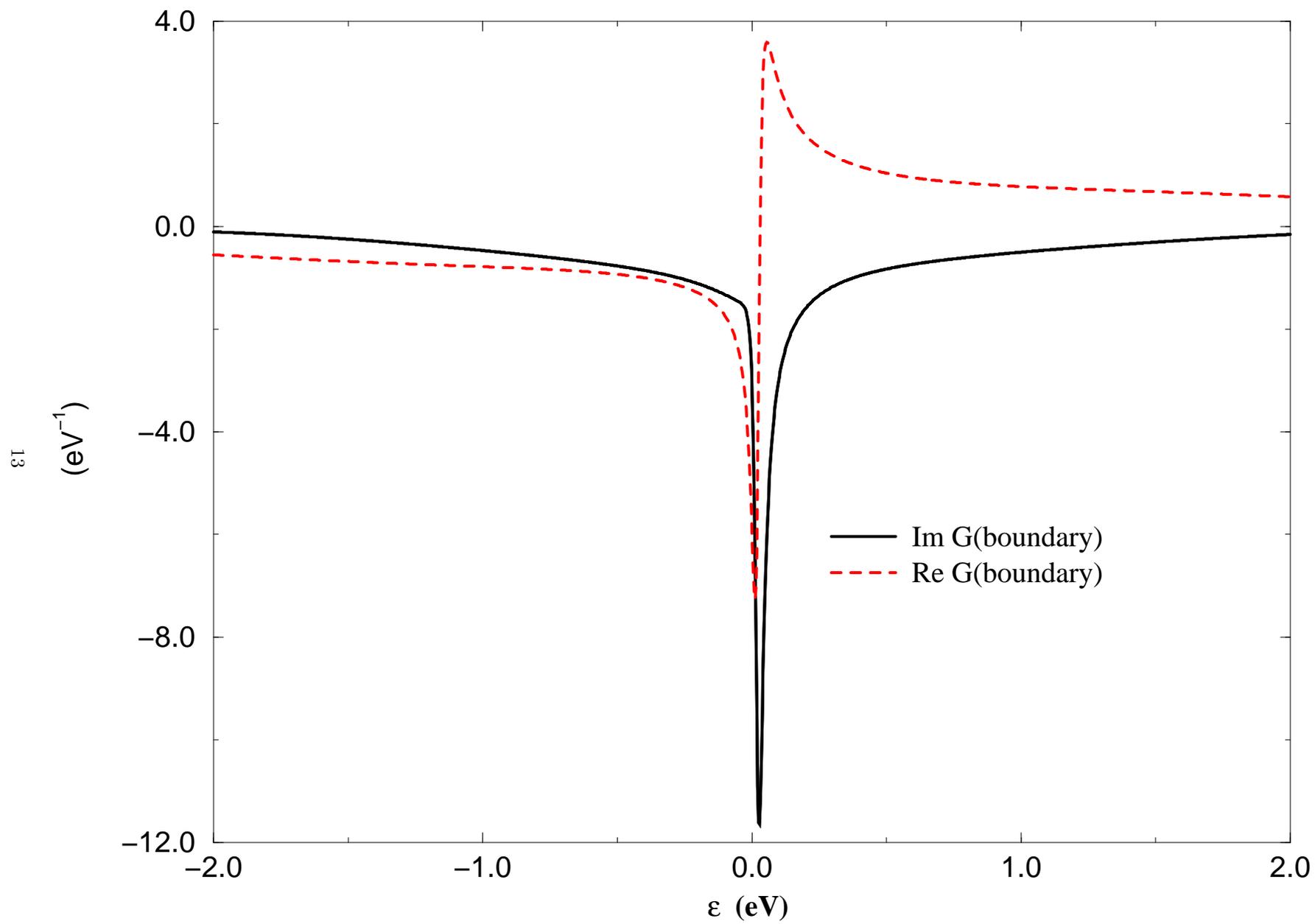
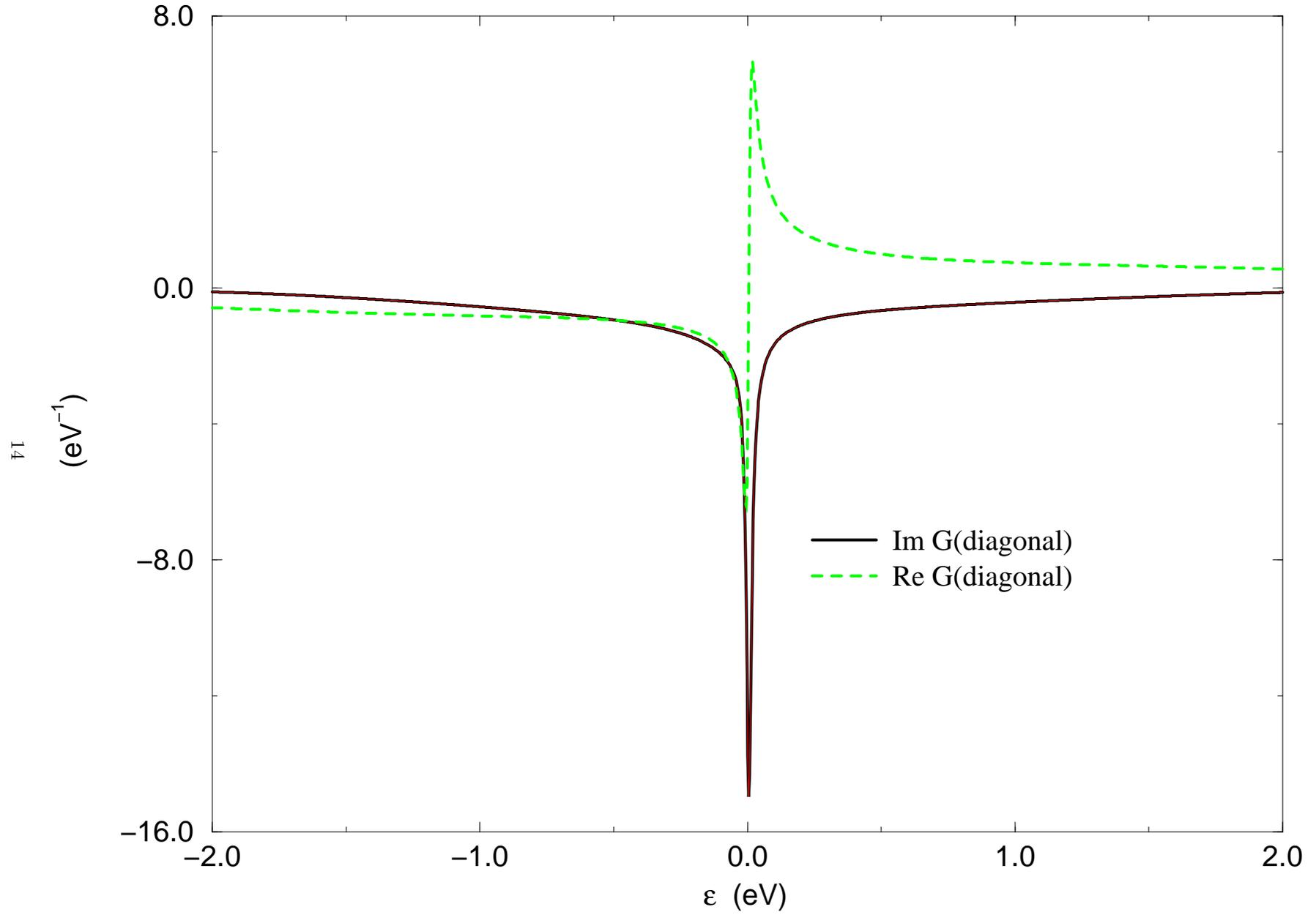
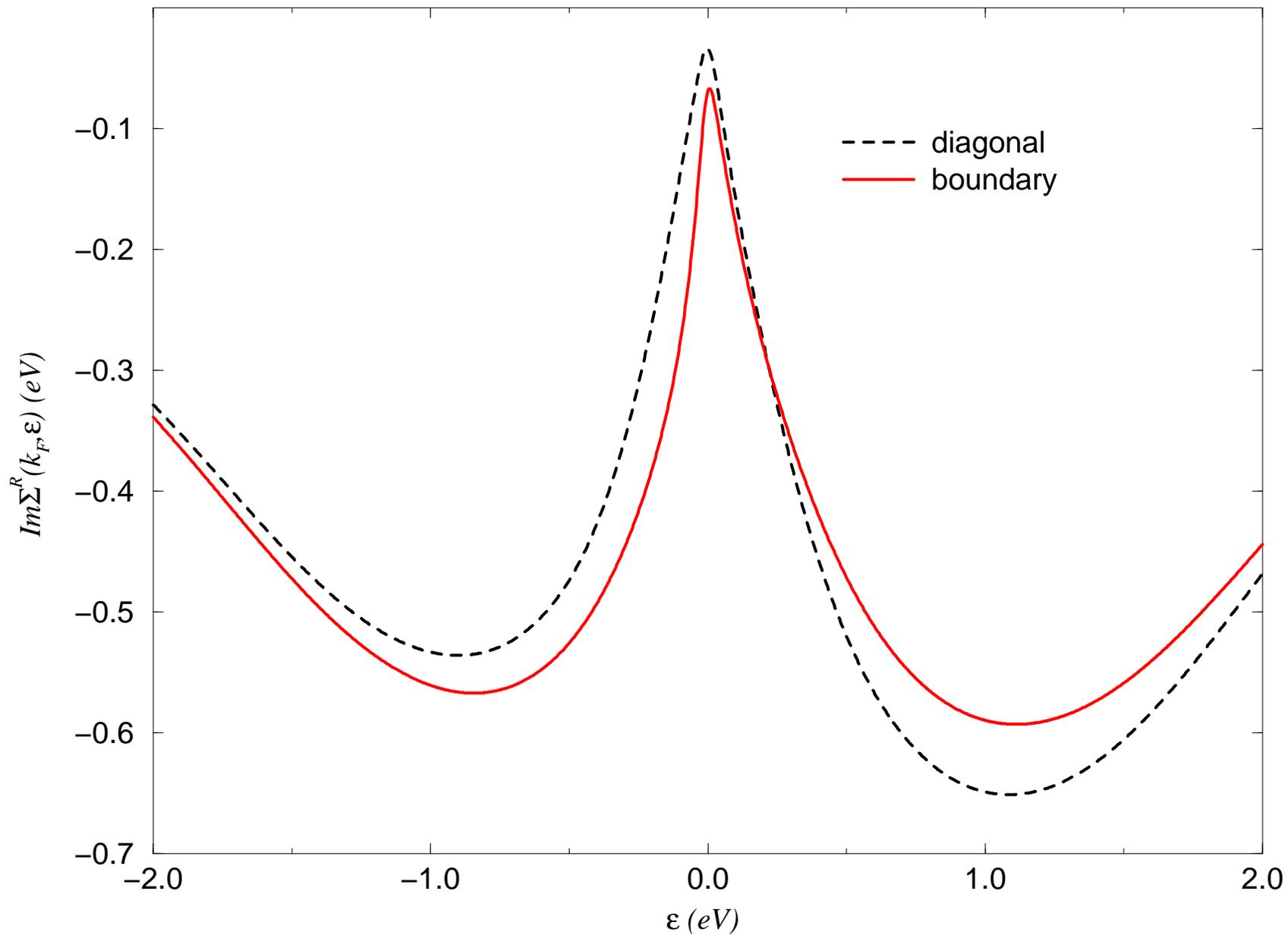
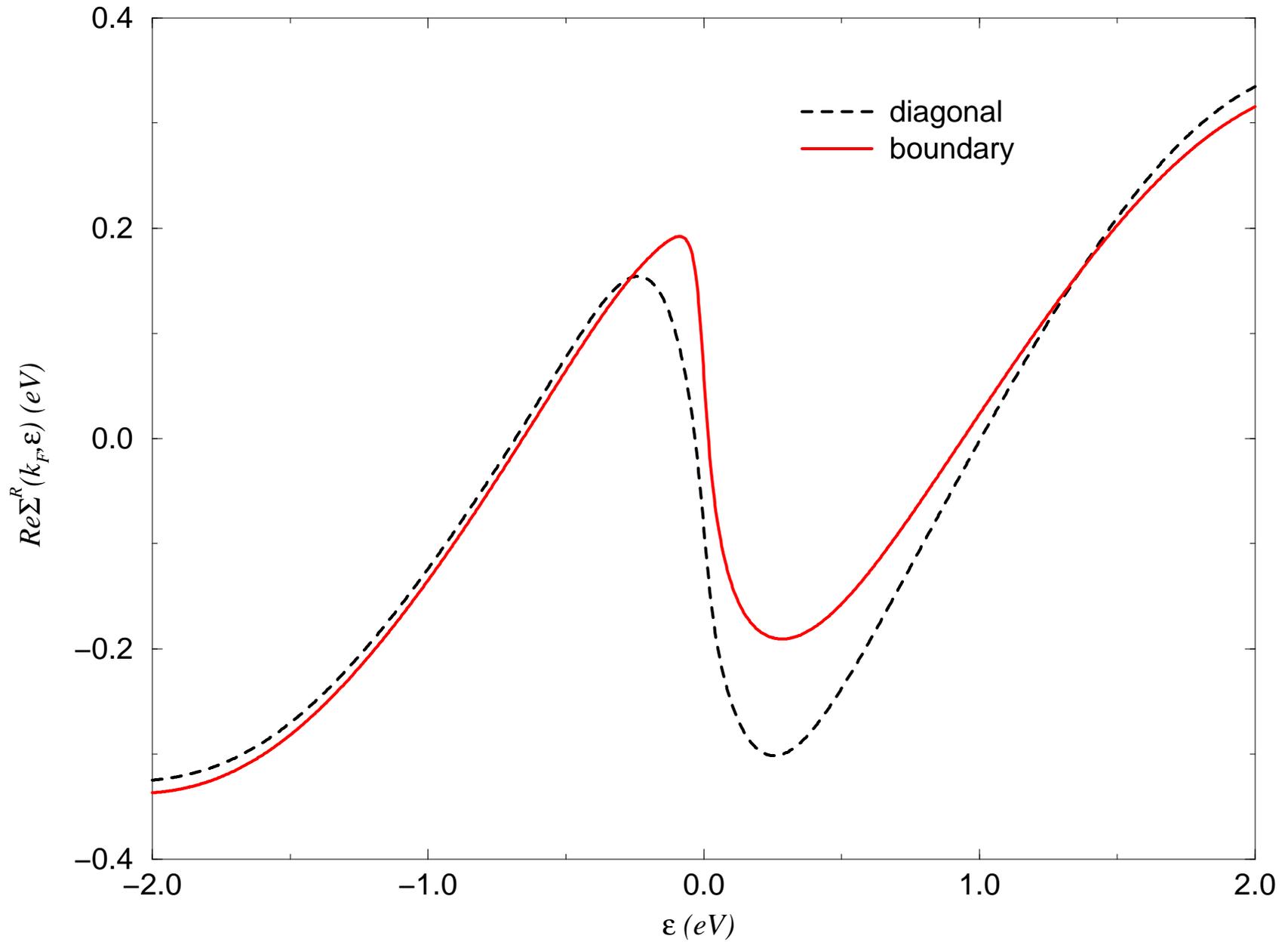
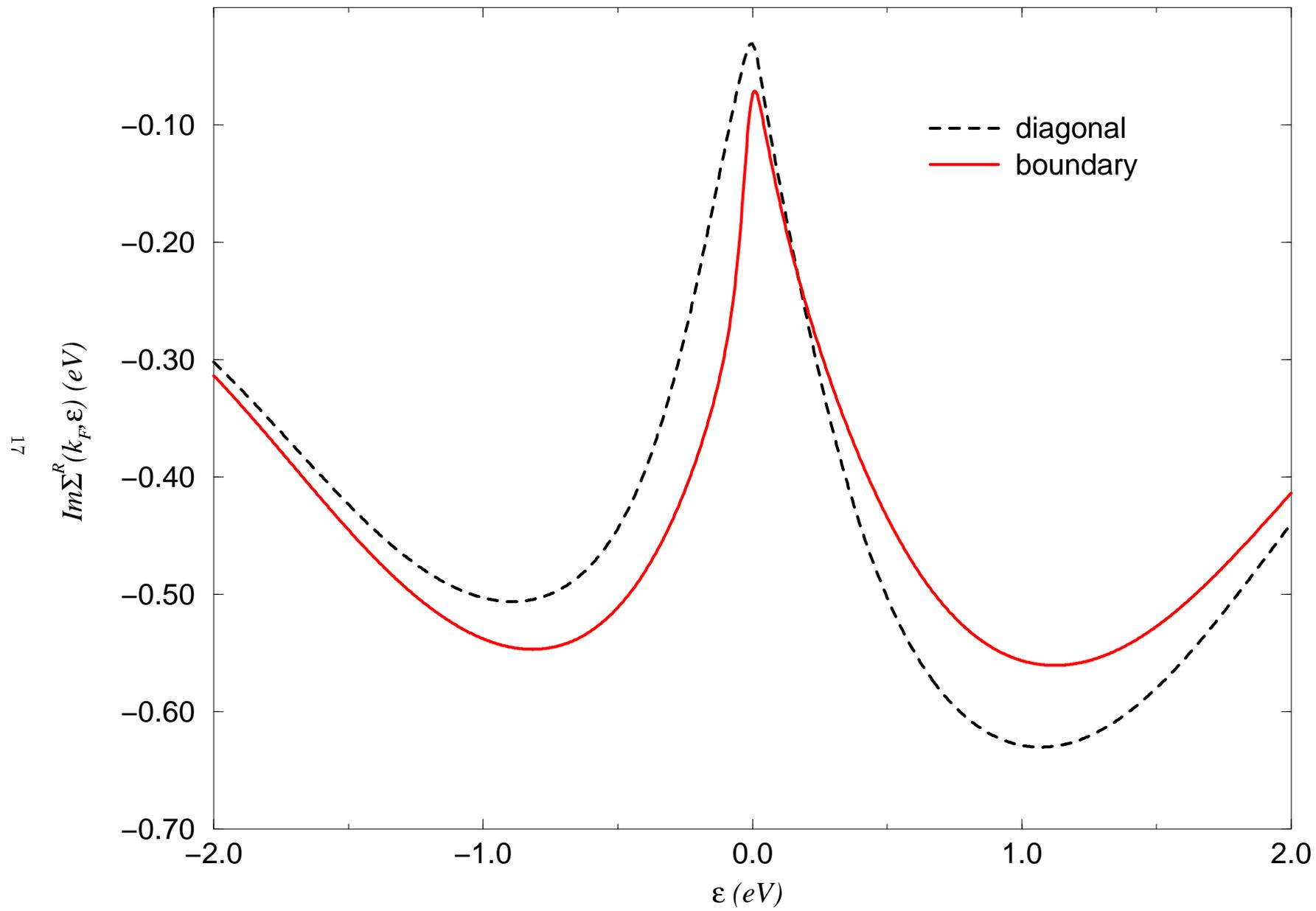
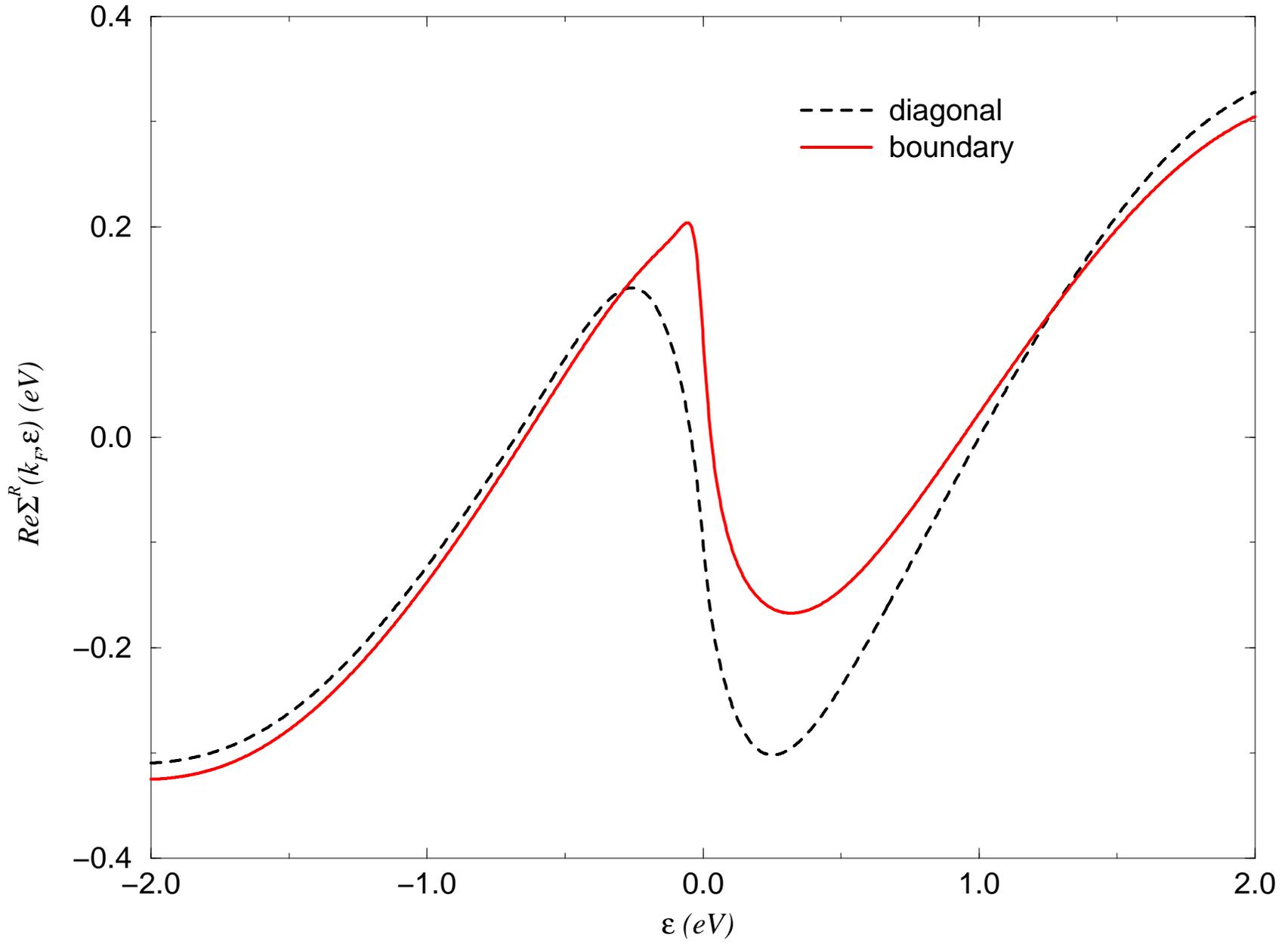
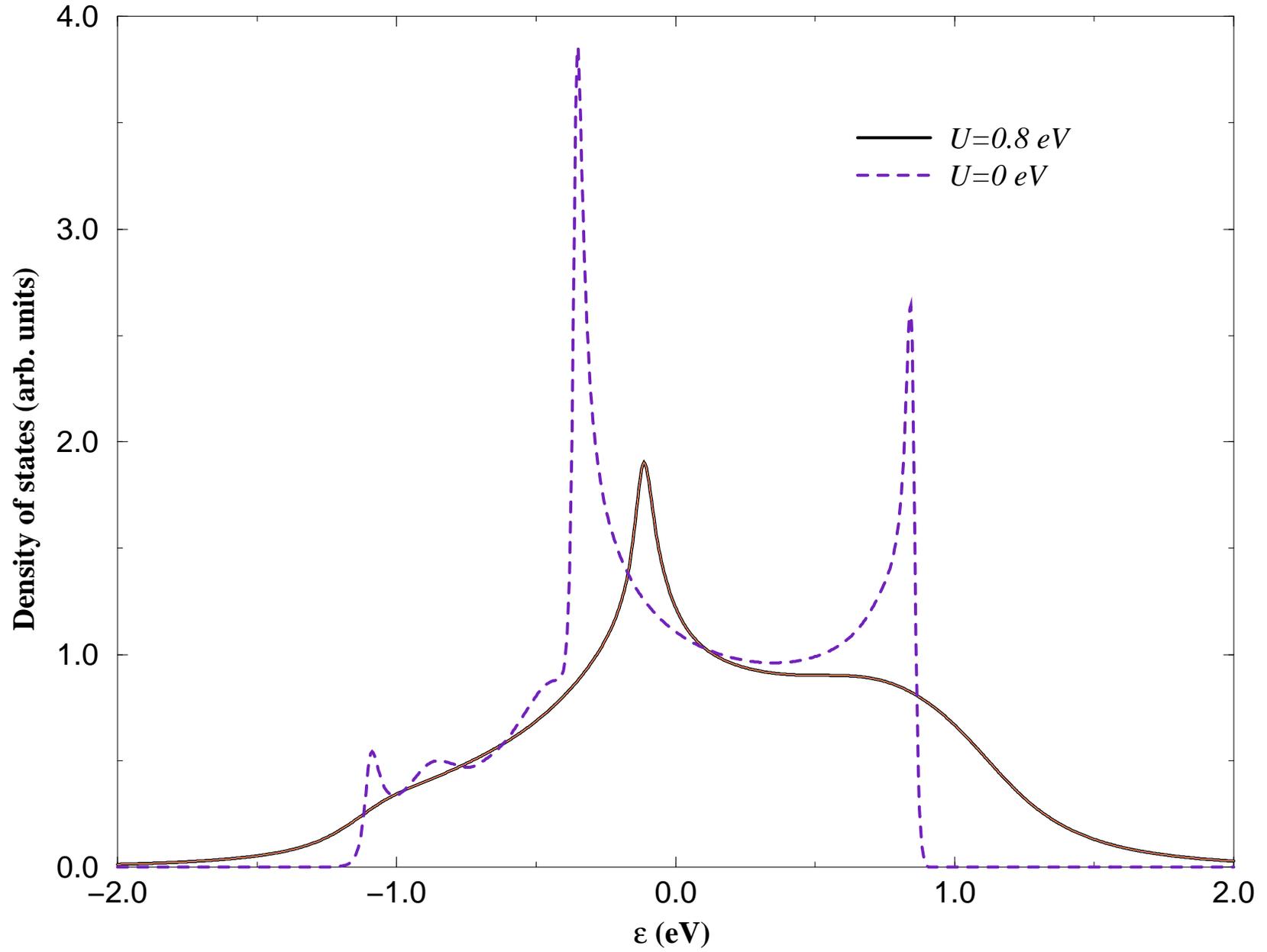
\end{document}